\documentclass[sigconf,authorversion,nonacm]{acmart}

\usepackage{float}
\usepackage{multirow}
\usepackage{colortbl}
\usepackage{enumitem}
\usepackage{color}
\usepackage{svg}
\usepackage{caption}
\usepackage{subcaption}
\usepackage{balance}

\AtBeginDocument{%
  \providecommand\BibTeX{{%
    \normalfont B\kern-0.5em{\scshape i\kern-0.25em b}\kern-0.8em\TeX}}}

\copyrightyear{2024}   
\acmYear{2024}
\setcopyright{acmlicensed}\acmConference[Websci '24]{ACM Web Science
Conference}{May 21--24, 2024}{Stuttgart, Germany}
\acmBooktitle{ACM Web Science Conference (WebSci '24), May 21--24, 2024, Stuttgart, Germany} 
\acmDOI{10.1145/3614419.3644024} 
\acmISBN{979-8-4007-0334-8/24/05} 

\acmConference[WebSci'24]{}{May 21--24,
  2024}{Stuttgart, Germany}

\begin{document}

\title{Shifting Climates: Climate Change Communication from YouTube to TikTok}

\author{Arianna Pera}
\email{arpe@itu.dk}
\affiliation{%
  \institution{IT University of Copenhagen}
  \streetaddress{Rued Langgaards Vej 7}
  \city{Copenhagen}
  \country{Denmark}
  \postcode{2300}
}

\author{Luca Maria Aiello}
\email{luai@itu.dk}
\affiliation{%
  \institution{IT University of Copenhagen}
  \streetaddress{Rued Langgaards Vej 7}
  \city{Copenhagen}
  \country{Denmark}
  \postcode{2300}
}

\renewcommand{\shortauthors}{Pera et al.}

\begin{abstract}
Public discourse on critical issues such as climate change is progressively shifting to social media platforms that prioritize short-form video content. Content creators acting on those platforms play a pivotal role in shaping the discourse, yet the dynamics of communication and audience reactions across platforms remain underexplored. To improve our understanding of this transition, we studied the video content produced by 21 prominent YouTube creators who have expanded their influence to TikTok as information disseminators. Using dictionary-based tools and BERT-based embeddings, we analyzed the transcripts of nearly 7k climate-related videos across both platforms and the 574k comments they received. We found that, when publishing on TikTok, creators use a more emotionally resonant, self-referential, and action-oriented language compared to YouTube. We also observed a strong semantic alignment between videos and comments, with creators who excel at diversifying their TikTok content from YouTube typically receiving responses that more closely align with their produced content. This suggests that tailored communication strategies hold greater promise in directing public discussion towards desired topics, which bears implications for the design of effective climate communication campaigns.
\end{abstract}



\maketitle

\section{Introduction}

Social media platforms have reshaped the landscape of public discourse, facilitating unprecedented global connectivity and empowering individuals to participate in critical societal debates, such as those surrounding climate change~\cite{pearce2019social}. The study of online social discourse is crucial for understanding interaction dynamics that may either impede or enhance processes of opinion formation~\cite{monti2022language}, democratic deliberation~\cite{jennings2021social}, and collective action~\cite{yasseri2016political}. Most prior studies in this area often have focused on traditional microblogging platforms such as Twitter or Reddit~\cite{pera2023measuring}, overlooking the continuous shifts of online audiences to new digital spaces~\cite{la2023drivers}, with younger crowds increasingly leaning towards platforms that emphasize video content, particularly in short form~\cite{stahl2023genz}.

Video-sharing platforms like YouTube and TikTok have emerged as powerful tools for disseminating information, shaping perceptions, and mobilizing communities~\cite{uldam2013online, hautea2021showing}. The dynamics on these platforms are largely driven by content creators, who play a critical role as both disseminators of information and opinion leaders and are fundamental actors in shaping a collective discourse on climate that can break out from escalating levels of polarization~\cite{falkenberg2022growing}. Nevertheless, the characteristics of climate communication on modern video-sharing platforms remain underexplored. In particular, one key open question is how content creators tailor their communication strategies to accommodate the substantial differences in content format and audience composition across different platforms~\cite{guinaudeau2022fifteen}, and how the public's response correlates to these changes. Enhancing our understanding of the interplay between storytelling and public reactions would provide valuable insights for refining strategies for climate communication online.

To improve our understanding of climate communication on video platforms, we conducted a comparative study of the YouTube and TikTok video productions of 21 prominent climate communicators active on both platforms. Crucially, by investigating the differences between the climate narratives of each creator across platforms, we contribute to the characterization of content adaptation strategies adopted by professional communicators, rather than merely describing the differences in climate-related content between YouTube and TikTok. To accomplish this, we use a text-analysis toolkit applied to video transcripts, descriptions, and user comments. We address two research questions:

\vspace{1pt} \noindent \textbf{RQ1.} \emph{What are the differences in the style and content of climate change communication between YouTube and Tiktok?} Drawing from previous qualitative literature, we focused on markers of emotions, individualism, and time focus, which are suggested to be typical elements of TikTok's style of interaction~\cite{hautea2021showing,barta2021constructing,stahl2023genz}.

\vspace{1pt} \noindent \textbf{RQ2.} \emph{Does the audience reaction semantically align with the reference content?} On video-sharing platforms, public discussions are centered around video content, giving content creators some control over the topics that receive public attention. We measured the semantic closeness between videos and their comments to quantify how much the creators' selection of topics defines the nature of the collective discourse on these platforms. Specifically, we investigated whether comments are semantically consistent with their corresponding videos (\textbf{RQ2.1}) and whether larger semantic shifts in a creator's content across platforms correlate with a greater semantic alignment of the public's reaction (\textbf{RQ2.2}).

\vspace{1pt} We found that TikTok narratives contain more calls to action, are more emotionally charged, and have a greater focus on the self rather than addressing others, compared to YouTube. Comments closely reflect video narratives in terms of semantics, and creators excelling at diversifying their TikTok content from YouTube tend to receive reactions that align more closely with their produced content. These insights can guide the design of future climate communication campaigns and inform quantitative studies on online discourse on video-sharing platforms.

\section{Methodology}

\subsection{Data collection}
We study content creators within four main categories of climate change communicators that have been researched consistently across prior studies in communication and environmental psychology: \emph{institutions} (including governments, NGOs, and politicians), \emph{news outlets}, \emph{science communicators}, and \emph{activists} (e.g, community-based movements and opinion leaders)~\cite{nerlich2010theory,schafer2012online}. Our decision to focus on creators rather than general climate change content is driven by our aim to characterize content adaptation strategies of users active on different platforms.

We began by identifying TikTok (TT) creators through a manual examination of the most popular accounts obtained via querying its Research API for ``climate change'' videos, analyzing the top 100 results returned. From each profile we deemed relevant to our scope, we also parsed the list of followed accounts, selecting those associated with climate change content. The full list was then manually categorized into the four communicator types and matched with the respective YouTube (YT) accounts based on usernames. The resulting set included 27 creators active on both platforms, later refined to 21 accounts who posted a minimum of three videos on each platform (Table~\ref{tab:accounts_types}).
\begin{table}
\footnotesize
 \caption{Content creators and their audience size.}
    \label{tab:accounts_types}
\begin{tabular}{lll}
\textbf{Type}                                                    & \textbf{Creators}                                                                                                                                                         & \textbf{tot. views}                                            \\ \hline
Institutions                                                     & \textit{UN climate change, MET office}                                                                                                                                    & \begin{tabular}[c]{@{}l@{}}TT: 39,6M\\ YT: 5,5M\end{tabular}   \\ 
News outlets                                                     & \textit{\begin{tabular}[c]{@{}l@{}}BBC news, The Guardian, DW Planet A.\\ Now this Earth\end{tabular}}                                                                    & \begin{tabular}[c]{@{}l@{}}TT: 24,6M\\ YT: 222,5M\end{tabular} \\ 
\begin{tabular}[c]{@{}l@{}}Science \\ communicators\end{tabular} & \textit{\begin{tabular}[c]{@{}l@{}}Climate Adam, Dr Gilbz, Our Changing Climate, \\ TED, Minute Earth, Zahra Biabani, Pique Action, \\ Margreen\end{tabular}}             & \begin{tabular}[c]{@{}l@{}}TT: 8,7M\\ YT: 76,7M\end{tabular}   \\ 
Activists                                                        & \textit{\begin{tabular}[c]{@{}l@{}}Extinction Rebellion, Greenpeace, Just Stop Oil, \\ Rupert Read, Friends of the Earth, Action for the \\ Climate, Parley\end{tabular}} & \begin{tabular}[c]{@{}l@{}}TT: 39,3M\\ YT: 83,2M\end{tabular}  \\ \hline
\end{tabular}
\end{table}
To isolate climate-related content from the video production of creators who cover diverse topics, we curated a list of keywords. To ensure both precision and high coverage, we started from a seed keyword and performed a keyword expansion step. We initiated this process on TT, sampling 100 videos per month matching the keyword ``climate change'' from September 2016 to November 2023. From the most frequently used hashtags in the retrieved videos, we identified 20 climate-related keywords\footnote{climate change, global warming, savetheplanet, climate crisis, greennewdeal, renewable energy, climateaction, gogreen, climatejustice, climatechange, globalwarming, climate\_change, savetheworld, climatecrisis, forclimate, savetheearth, stopclimatechange, renewableenergy, climatechangeisreal, saveourplanet}. We then used this list to query the TT Research API and the YT API. 
In total, we collected 2,617 TT videos and 4,286 YT videos. Table~\ref{tab:accounts_types} reports an overview of the audience of climate change communication by creator type. The total number of views is 112,2M on TT and 387,8M on YT. We highlight that both regular videos and YT's \emph{shorts} are included in the YT set; currently, it is not feasible to directly filter for \emph{shorts} through the YouTube API. Thus, the distribution of YT video lengths is varied, as shown in Table~\ref{tab:quantiles}.

\begin{table}[t!]
\centering
\footnotesize
\caption{Quantiles of YT video duration.}
\label{tab:quantiles}
\begin{tabular}{c|c|c|c|c|c}
\textbf{quantile} & $q_{25}$ & $q_{50}$ & $q_{75}$  & $q_{90}$ \\ \hline
\textbf{length}  & 0m58s   & 3m36s & 10m15s & 26m57s 
\end{tabular}
\end{table}

For the analysis of video content, we relied on two data sources: transcripts and video descriptions. Transcripts were available for a subset of auto-captioned videos on both platforms. On YT, when auto-captions were not available, we employed the Whisper audio-to-text converter to retrieve missing transcripts~\cite{radford2023robust}. However, this conversion could not be applied to TT videos without infringing the Terms of Service. In instances where transcripts were unavailable (7.7\% of YT videos and 79.8\% of TT videos), we used video descriptions as an alternative. Given the high cosine similarity between Sentence-BERT~\cite{reimers2019sentence} embedding representations of transcripts and descriptions in videos that contain both (0.471 on TT and 0.463 on YT), we deemed descriptions as suitable proxies for the video content. For simplicity, throughout the paper, we use the term \emph{transcript} to refer to any textual description of a video.

We conducted minimal pre-processing on the textual data, excluding transcripts with fewer than five unique words and removing text within music tags indicating musical pieces within a YT transcript. Descriptions underwent cleaning, including the removal of website links, mentions, hashtags, references to other platforms, and recurring expressions, such as account descriptions. 
\begin{table}[t!]
\centering
\footnotesize
\caption{Quantiles of comments descriptive variables.}
\label{tab:quantiles_comments}
\begin{tabular}{c|c|c|c|c|c}
\multicolumn{1}{l|}{}                                                            & \textbf{platform} & \textbf{q25} & \textbf{q50} & \textbf{q75} & \textbf{q90} \\ \cline{1-6} 
\multirow{2}{*}{\textbf{\begin{tabular}[c]{@{}c@{}}comments\\per video\end{tabular}}} & \textit{TT}       & 3            & 9            & 29           & 100          \\ \cline{2-6} 
                                                                                 & \textit{YT}       & 2            & 15           & 81           & 383          \\ \hline
\multirow{2}{*}{\textbf{\begin{tabular}[c]{@{}c@{}}tokens per\\ comment\end{tabular}}} & \textit{TT}       & 1            & 7            & 15           & 23           \\ \cline{2-6} 
                                                                                 & \textit{YT}       & 7            & 15           & 32           & 64          
\end{tabular}
\end{table}

To study the reaction to climate change communication, we retrieved comments to creators' videos. In particular, we called the \texttt{CommentThreads} method in the YT API to collect up to 200 comments and first-level replies in YT and the \texttt{VideoComments} method in the TT API to collect up to 100 comments and nested replies. We collected a total of 81,750 comments for TT and 492,788 comments for YT. Table \ref{tab:quantiles_comments} reports the quantiles for the number of comments per video and the number of tokens per comment, for TT and YT.

\subsection{Metrics}

\paragraph{\textbf{Popularity}}

We quantify the popularity of content creators on TT relative to YT with the ratio between the average number of views per video published on the two platforms:
\begin{equation}
    \text{\emph{relative popularity(u)}} = \frac{avg(\text{views}^u_{TT})}
    {avg(\text{views}^u_{YT})}
\label{eq:relative_popularity}
\end{equation}
Values considerably deviating from 1 indicate a creator being more popular on TT ($\gg 1$) or YT ($\ll 1$).

\begin{figure}[t!]
    \centering   
    \includegraphics[width=0.45\textwidth]{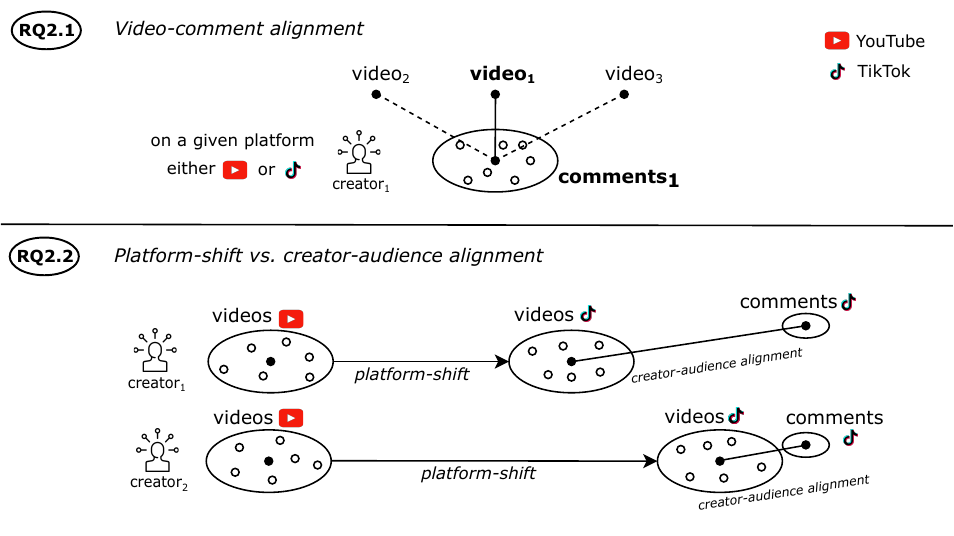}
    \caption{Alignment of comments to corresponding videos (\textbf{RQ2.1}) and overall content-reaction alignment related to platform stylistic shift (\textbf{RQ2.2}).}
    \label{fig:schema}
\end{figure}

\paragraph{\textbf{Language metrics}} 

We employed established linguistic tools to conduct video transcript analysis. The NRC lexicon~\cite{mohammad2013crowdsourcing} gauged the proportion of words expressing \emph{emotions} (anger, disgust, fear, joy, sadness, and surprise). The LIWC psycho-linguistic dictionary identified the fraction of words related to \emph{sentiment} (positive or negative), \emph{temporal focus} (present, past, or future), and \emph{personal pronouns} usage. We computed an \emph{emotionality} measure as the mean relative frequency of emotional words and a \emph{sentiment} measure as the difference between positive and negative word frequencies. Considering the important role of personal pronouns in capturing personal identity from text~\cite{orvell2022you}, we derived an \emph{individualism index} from pronoun fractions:
\begin{equation}
\text{\emph{individualism}} = 0.5 + \frac{0.5 \cdot (\text{\emph{f}}_{\text{I}} - \text{\emph{{f}}}_{\text{we}} - \text{\emph{{f}}}_{\text{you}} - \text{\emph{{f}}}_{\text{he|she|they}})}{\text{\emph{f}}_{\text{I}} + \text{\emph{{f}}}_{\text{we}} + \text{\emph{{f}}}_{\text{you}} + \text{\emph{{f}}}_{\text{he|she|they}} + 1}
\label{eq:individualism}
\end{equation}
where $\emph{f}_{\bullet}$ denotes the relative frequency of a pronoun group in the text. Values around 0.5 suggest a balance between the use of first-person pronouns and other personal pronouns, while values closer to 0 or 1 indicate a focus on the community or the self, respectively. This measurement represents a best-effort approach to establish a unified metric that can capture the equilibrium between self-focus and attention to consideration of others. Yet, we acknowledge that the use of one single score may offer an oversimplified perspective on the nuanced relationship between identity and pronouns.

\paragraph{\textbf{Semantic alignment}} 

We used Sentence-BERT~\cite{reimers2019sentence} to convert text into embeddings for semantic comparison, considering three similarity measures to assess the alignment of video-comment pairs and across video corpora on different platforms.
For a video $v_i$ posted by a creator $u$ and its $n$ comments $C_i = {c_{i1}, c_{i2}, ..., c_{in}}$, let $\textbf{v}_i$ represent the Sentence-BERT embeddings vector of the video transcript and $\overline{\textbf{C}_i}$ represent the average embedding vector of all its comments. We quantified the similarity between the two by computing the difference between the video-comment similarity and the average similarity that the comment vector has with each of the other video vectors from the same creator, given a total of $m$ posted videos:
\begin{equation}
    \text{\emph{video-comment alignment}}(v_i) = cos({\overline{\textbf{C}_i}}, \textbf{v}_i) - \frac{\sum_{j=1, j \neq i}^{m} cos({\overline{\textbf{C}_i}}, \textbf{v}_j)} {(m-1)}
\label{eq:video-comment-alignment}
\end{equation}
Values exceeding 0 indicate comments being semantically closer to the video they reference than to the content of other videos from the same creator. We used this measure to characterize creators by averaging it over all the videos they posted on either platform. 

To determine how closely a creator $u$'s narrative aligns with their audience's reaction, we calculated the similarity between the average embedding of the videos they posted ($\overline{{\textbf{V}^{u}}}$) and the average embedding of the comments that all their videos received ($\overline{{\textbf{C}^{u}}}$):
\begin{equation}
    \text{\emph{creator-audience alignment}} = cos(\overline{{\textbf{V}^{u}}}, \overline{{\textbf{C}^{u}}})
    \label{eq:centroid-video-comment}
\end{equation}
Finally, to assess the semantic shift between a creator's YT and TT productions, we calculated the cosine similarity between the average embeddings of videos on the two platforms:
\begin{equation}
    \text{\emph{platform-shift}} = cos(\overline{{\textbf{V}^{u}_{YT}}}, \overline{{\textbf{V}^{u}_{TT}}})
    \label{eq:platform-shift}
\end{equation}
These three measurements enabled us to test whether the audience discourse significantly aligns with the narrative of the corresponding video content (\textbf{RQ2.1}, Eq.~\ref{eq:video-comment-alignment}), and whether the magnitude of semantic shifts in a creator's narration, particularly between one platform and another (Eq.~\ref{eq:platform-shift}), is associated with the degree of alignment between such overall narrative and the public's reaction (\textbf{RQ2.2}, Eq.~\ref{eq:centroid-video-comment}). This concept is illustrated in Figure~\ref{fig:schema}.

\section{Results}

\subsection{Types of climate change communication}

\begin{figure}[t!]
    \centering
    \includegraphics[width=0.3\textwidth]{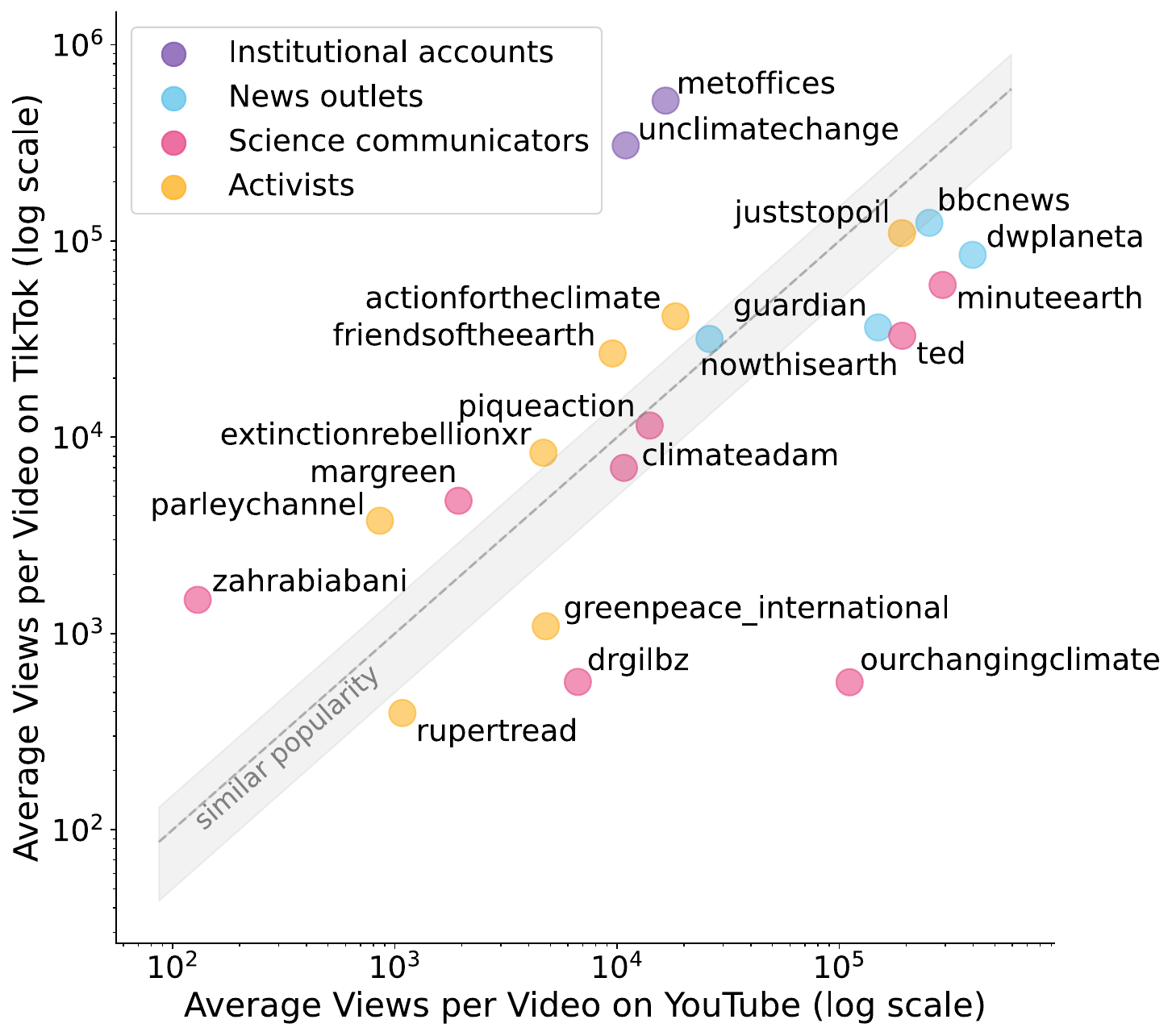}
    \caption{Popularity of creators on YT and TT.}
    \label{fig:accounts_popularity}
\end{figure}

All climate change communicators in our study, with the sole exception of Margreen, started using YT as a platform for content dissemination before transitioning to TT --- on average, 5.8 years before their TT debut. Creators exhibit varying degrees of popularity, with some enjoying greater favor on TT (e.g., \emph{unclimatechange, metoffices}), others finding more success on YT (e.g., \emph{ourchangingclimate}), and a few achieving a balance between the two platforms, as illustrated in Figure \ref{fig:accounts_popularity}. 

\begin{figure}[t!]
    \centering
    \includegraphics[width=0.47\textwidth]{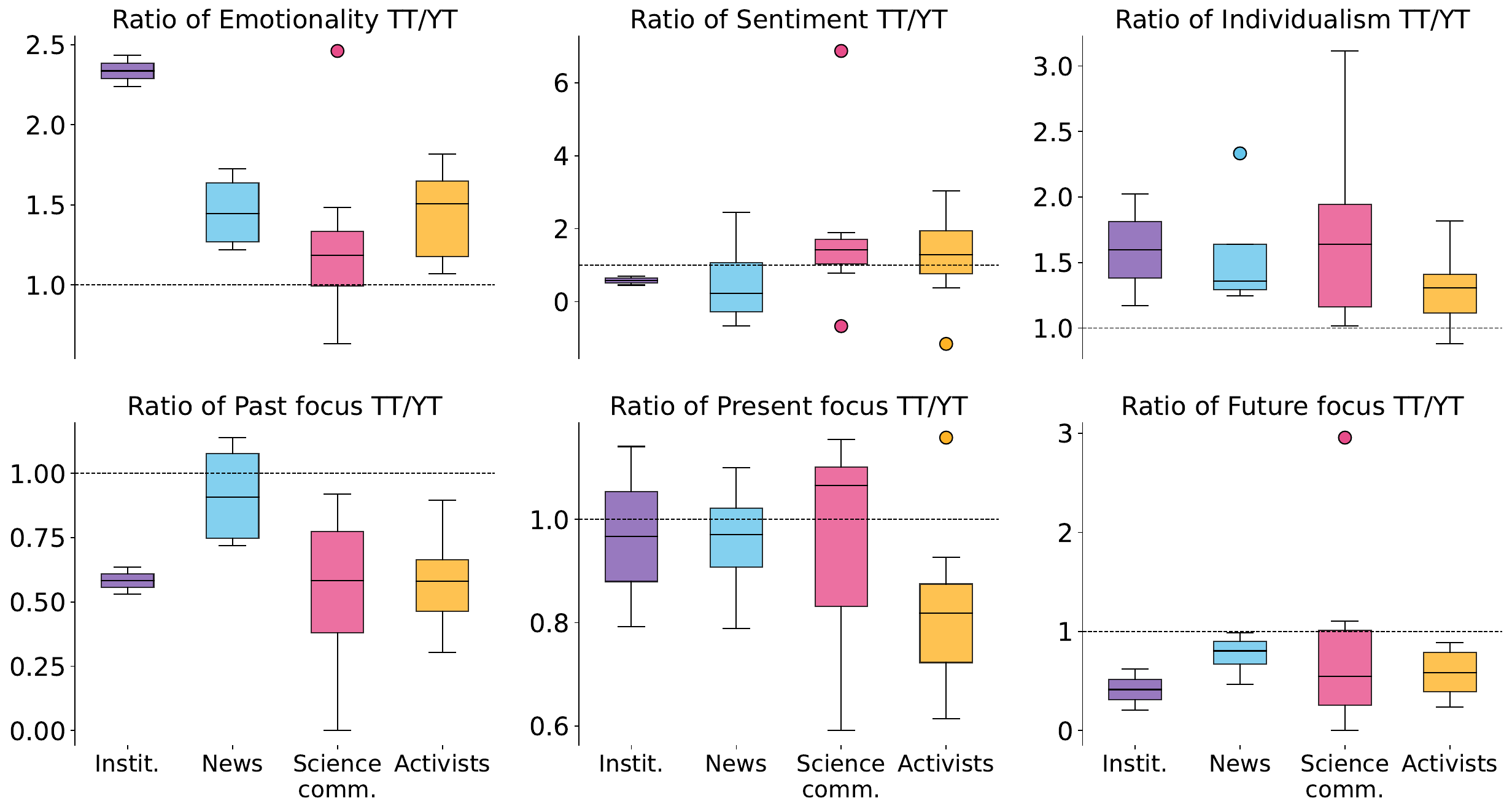}
    \caption{Distributions of ratios between linguistic features found in the TT and YT videos of each creator, grouped by creator category.}
    \label{fig:diff_lang_creator}
\end{figure}

To assess the diversity of communication between YT and TT, we started by analyzing linguistic cues used in videos (Figure~\ref{fig:diff_lang_creator}). TT videos reference emotions more frequently and offer more self-focused narratives than those on YT. This aligns with qualitative studies on TT, which emphasize the prevalence of self-reflexivity and self-representation, particularly among younger contributors~\cite{hautea2021showing,stahl2023genz}, and the frequent use of highly emotional language~\cite{barta2021constructing}. Narratives on YT, on the other hand, tend to reference past or future events more frequently compared to TT. These linguistic styles are generally consistent across all types of content creators, with a few notable exceptions. For instance, science communicators and activists typically express more positive sentiment on TT than on YT, while institutions and news outlets tend to convey more negative sentiments. The difference between the style used in YT and TT is less predictable for science communicators compared to other categories across all linguistic measures, suggesting a wide range of communication styles within this category.

\begin{figure}[t!]
    \centering
    \includegraphics[width=0.3\textwidth]{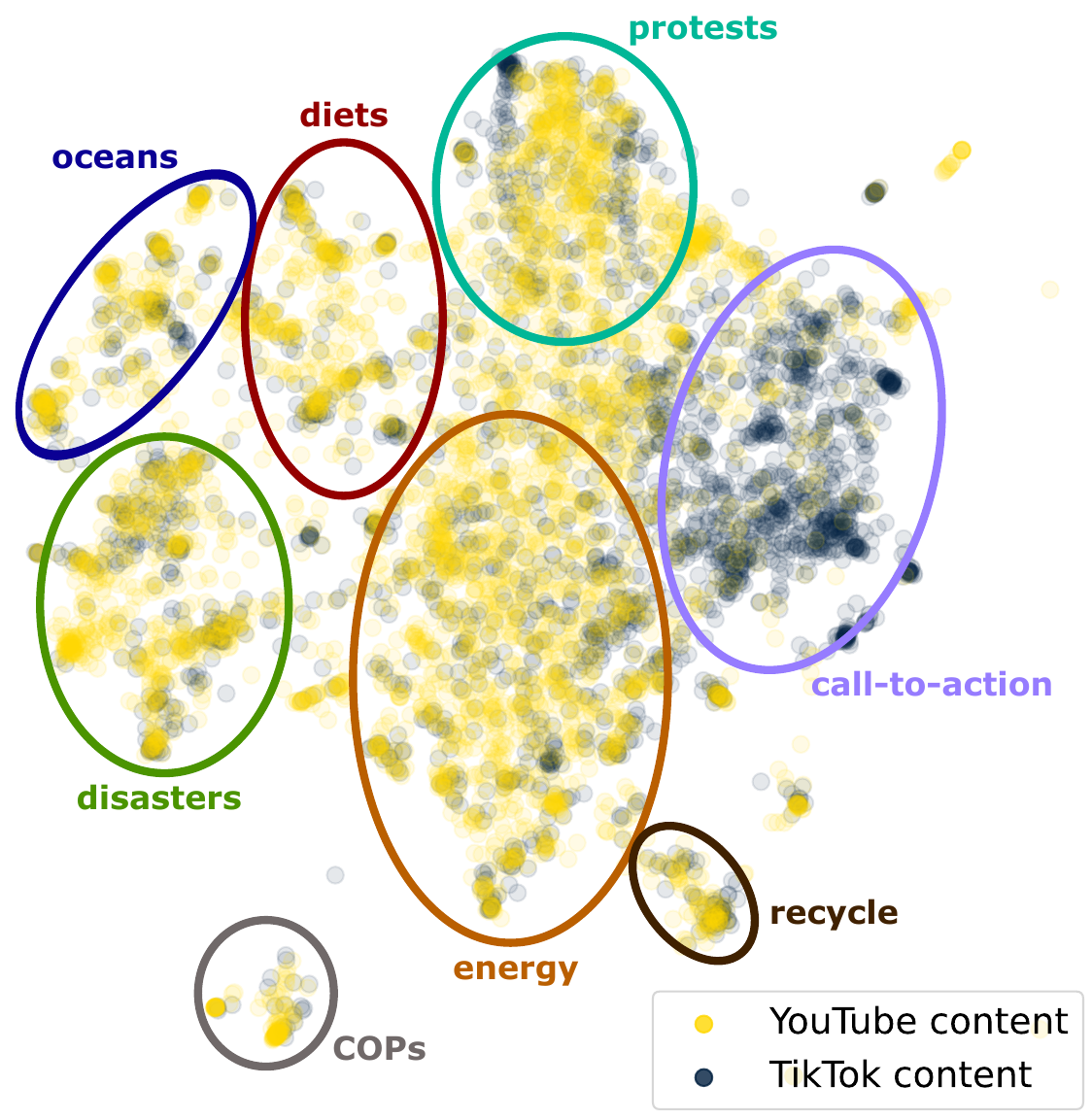}
    \caption{UMAP representation of YT and TT content, with annotated topics.}
    \label{fig:umap_annotated}
\end{figure}

To determine whether the differences in linguistic tone across the two platforms can be attributed to differing topical focuses, we mapped all embedding vectors of video transcripts from the two platforms onto a 2-dimensional projection using UMAP~\cite{mcinnes2018umap}. Subsequently, we applied HDBSCAN~\cite{campello2013density} to the stacked UMAP embeddings of YT and TT content, resulting in the identification of 37 clusters. To enhance clarity, we merged smaller clusters that were closely situated within the same density region. This consolidation yielded a topical map of 8 distinctive clusters, as depicted in Figure~\ref{fig:umap_annotated}. Upon inspecting a random selection of 20 transcripts from each cluster, we found they represent distinct topics (Table~\ref{tab:example_topics}): protest reports, petition signing and group joining campaigns, plant-based diet and environmental protection, oceans at risk, natural disasters, energy and emissions, recycling, and Conferences Of the Parties (COPs). Notably, TT content exhibited a pronounced concentration on topics related to calls to action. This observation aligns with the interactive practices attributed to TT, as highlighted in social science studies~\cite{abidin2021mapping}, and with the psychological impact of communicating emotions on decision-making processes~\cite{achar2016we}, emphasizing an association between emotionality and call-to-action events on TT.

\begin{table}[]
\scriptsize
\centering
\caption{Examples of content by topic.}
\label{tab:example_topics}
\begin{tabular}{ccll}
\textbf{Topic}                                                                                                              &  & \multicolumn{1}{c}{\textbf{Content example}}                                                                     &           \\ \cline{1-3}
\multicolumn{1}{c|}{\multirow{2}{*}{\textbf{\begin{tabular}[c]{@{}l@{}}Call-to\\action\end{tabular}}}}                                                              & YT           & 
\textit{...Let’s get Ohio leaders to pass the Energy Jobs and Justice Act now...}                               &           \\ 
\multicolumn{1}{c|}{}                                                                                                       & TT            & 
\textit{Join us in demanding your governor declare a climate emergency...}                                    & \textit{} \\ \cline{1-3}
\multicolumn{1}{c|}{\multirow{2}{*}{\textbf{COPs}}}                                                                         & YT           & 
\textit{The UN Climate Change Conference (COP 26) in Glasgow, United Kingdom...}                              &           \\ 
\multicolumn{1}{c|}{}                                                                                                       & TT            & 
\textit{The UN Climate Conference COP 27 in Egypt started today...}                                             &           \\ \cline{1-3}
\multicolumn{1}{c|}{\multirow{2}{*}{\textbf{Protests}}}                                                             & YT           & 
\textit{...welcome to Sunday's weekly roundup with this rebellion...}
                     &           \\ 
\multicolumn{1}{c|}{}                                                                                                       & TT          &  
\textit{I'm here in the road today because things are desperate...}
                  &           \\ \cline{1-3}
\multicolumn{1}{c|}{\multirow{2}{*}{\textbf{Recycle}}}                                                                    & YT           & 
\textit{Plastics have been in production for less than a hundred years...}
       &           \\ 
\multicolumn{1}{c|}{}                                                                                                       & TT            & 
\textit{Reminder: plastic doesn't break down it breaks up!}
                    &           \\ \cline{1-3}
\multicolumn{1}{c|}{\multirow{2}{*}{\textbf{\begin{tabular}[c]{@{}c@{}}Energy\end{tabular}}}}               & YT           & 
\textit{These men are some of the biggest polluters on the planet...}
                &           \\ 
\multicolumn{1}{c|}{}                                                                                                       & TT            & 
\textit{How can the UK secure affordable energy for everyone?...}
                                 &           \\ \cline{1-3}
\multicolumn{1}{c|}{\multirow{2}{*}{\textbf{\begin{tabular}[c]{@{}c@{}}Diet\end{tabular}}}} & YT           & 
\textit{We really looked at companies that were adopting plant-based...}
               &           \\
\multicolumn{1}{c|}{}                                                                                                       & TT            & 
\textit{Did you know that pig production in just the top 4 factory farming...}
 &           \\ \cline{1-3}
\multicolumn{1}{c|}{\multirow{2}{*}{\textbf{Disasters}}}                                                            & YT           & 
\textit{...workers who keep flights running during a winter storm...}
                        &           \\ 
\multicolumn{1}{c|}{}                                                                                                       & TT            & 
\textit{It's been freaking hot this week we've actually just had some of the hottest...}
             &           \\ \cline{1-3}
\multicolumn{1}{c|}{\multirow{2}{*}{\textbf{Oceans}}}                                                                       & YT           & 
\textit{Underneath the rocking blues of the ocean’s surface lie forests of green...}
        &           \\ 
\multicolumn{1}{c|}{}                                                                                                       & TT            & 
\textit{More than 600 tons of dead marine washed ashore in Florida due to red tide.}                           &       \\ \cline{1-3}   
\end{tabular}
\end{table}

\begin{figure*}
     \centering
     \begin{subfigure}[b]{0.42\textwidth}
         \centering
         \includegraphics[width=0.78\textwidth]{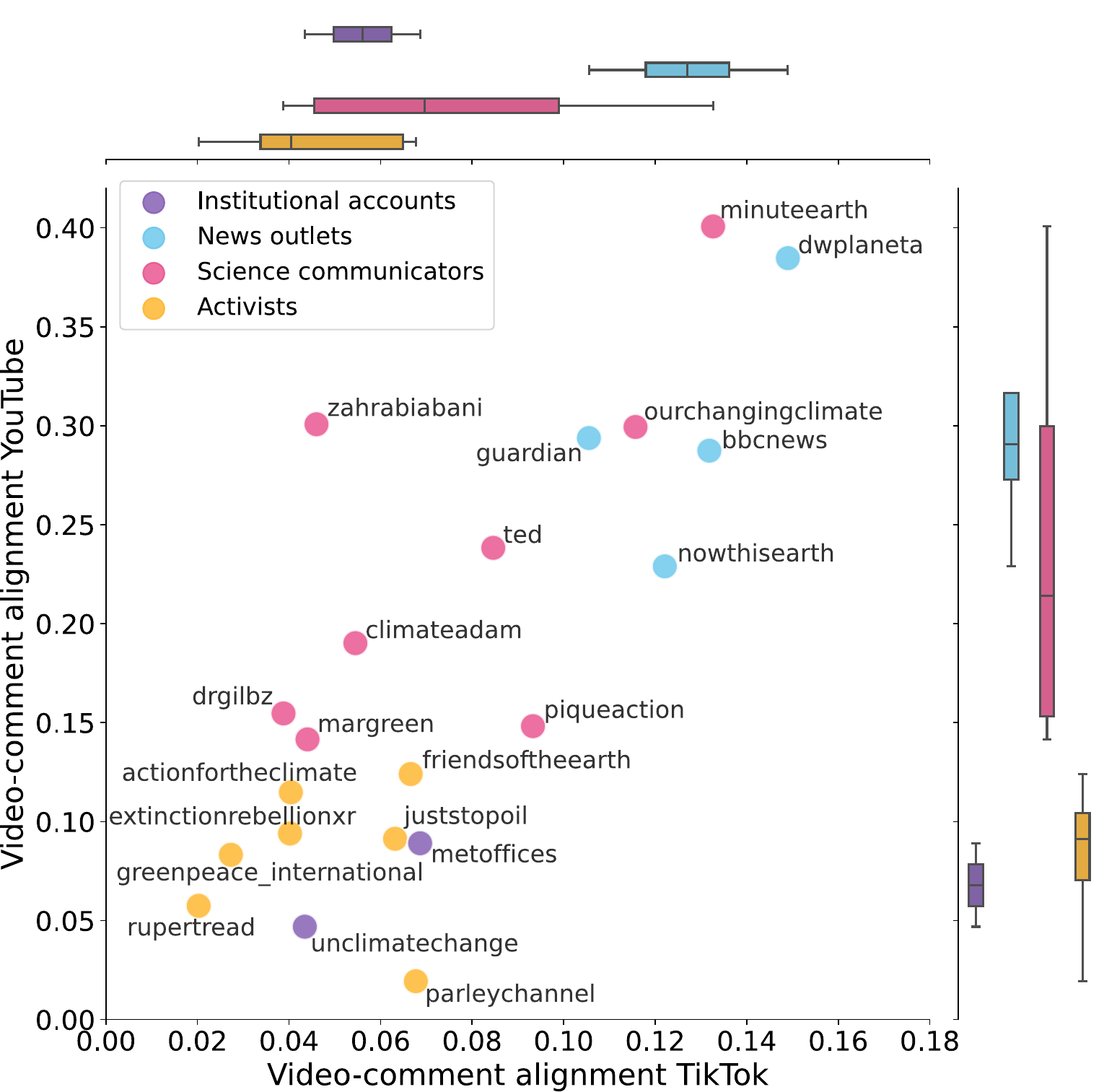}
         \caption{\emph{Video-content alignment} for YT and TT.
         }
\label{fig:reactions_videos_creator}
     \end{subfigure}
     \hspace{5mm}
     \begin{subfigure}[b]{0.42\textwidth}
         \centering
         \includegraphics[width=0.78\textwidth]{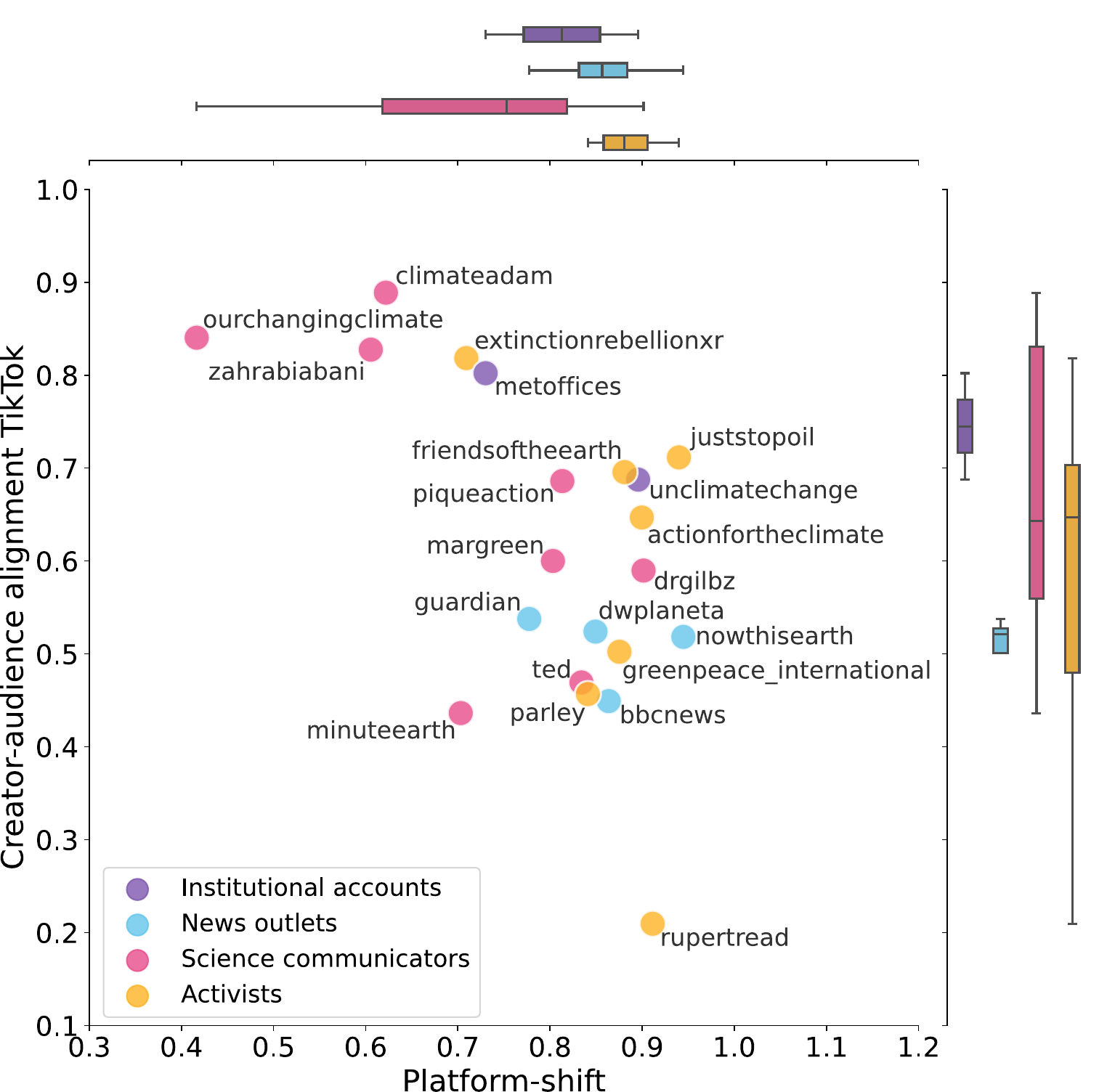}
         \caption{\emph{Platform-shift} vs. \emph{creator-audience alignment} on TT.}\label{fig:delta_comments_videos_TT_creator}
     \end{subfigure}
        \caption{Semantic conformity in terms of alignment of comments to their reference content (a) and alignment of TT audience to TT creators given the diversity of content from YT (b). Boxplots show distributions of the values along the two axes.}
        \label{fig:sem_align}
\end{figure*}

\subsection{Semantic shift and comment alignment}
To evaluate the alignment between content and reactions, we first assessed whether the narratives chosen by creators have an impact on the nature of the collective discourse that emerges from users commenting on their videos. We estimated the degree of semantic alignment between videos and their respective comment threads using the \emph{video-comment alignment} measure (Eq.~\ref{eq:video-comment-alignment}). Across all creators and both platforms, comments generally align with the semantics of the videos they reference (Figure~\ref{fig:reactions_videos_creator}), especially among science communicators and news outlets. Specifically, we observed that \emph{video-comment alignment} tends to be higher on YouTube, with values ranging from 0.019 to 0.385, compared to TikTok, where values range from 0.020 to 0.149. To explore the potential confounding effect introduced by platform choice, we computed the cosine similarity between a randomly selected set of 100 videos and their corresponding comment centroids for both YouTube and TikTok. The confidence interval of the average of this similarity measure is $0.349 \pm 0.016$ for YouTube and $0.246 \pm 0.036$ for TikTok. As a result, while comments typically demonstrate a stronger alignment with videos on YouTube in contrast to TikTok, such alignment is even stronger for the YouTube videos considered in this study.
This finding quantifies creators' influence not only in reaching broad audiences but also in guiding online discussions towards specific themes. While our study is purely observational and doesn't conclusively establish a causal link between video content and public reaction, the content-centric nature of video-sharing platforms grants creators some control over emerging topics in public discourse.

Lastly, we investigated whether a greater degree of content adaptation to the platform is associated with a stronger semantic alignment between video and comments. In our specific context, most content creators began their careers on YT, and the degree to which they adapt their TT content to appeal to a new audience may correlate with how closely the audience can relate to that content. We found an inverse correlation (Pearson $r=-0.569$, $p < 0.01$) between the creators' \emph{platform-shift} (i.e., the semantic similarity between their YT and TT content) and their \emph{creator-audience alignment} on TT (i.e., the semantic similarity between videos and comments on TT, defined in Eq.~\ref{eq:centroid-video-comment}). This suggests that creators who diversify their content the most when transitioning from YT to TT also receive TT reactions that most closely align with their proposed narrative.

\section{Discussion and conclusion} \label{sec:conclusion}
Climate change stands out as one of the key social dilemmas of our century. Video-sharing platforms like YouTube and TikTok offer ideal avenues for dynamic communication on this pressing topic. TikTok, in particular, has established itself as an exceptionally dynamic platform, emphasizing the preference for concise communication. Content creators transitioning from YouTube to TikTok might then need to recalibrate their strategies to align with the unique dynamics of this platform. 

Our analysis of the video content of 21 climate change communicators active on both YouTube and TikTok uncovered some of these adaptation mechanisms. When transitioning to TikTok, seasoned YouTube creators tend to adopt more emotional and self-centered narratives when addressing climate change, with the predominant topic being calls to action (\textbf{RQ1}). The reactions to climate change narratives are generally aligned with the corresponding content, especially among science communicators and news outlets (\textbf{RQ2.1}). Notably, creators excelling in diversifying TikTok content from YouTube also tend to receive reactions that align more closely with their content, which provides initial evidence of the potential of some creators to steer the public discussion towards desired topics (\textbf{RQ2.2}). 
Our findings highlight implications for analysts and interventions, emphasizing the need to recognize platform distinctions in content diversity and the importance of tailoring content for aligned audience reactions. 

While we aimed to conduct a fair comparison between YouTube and TikTok by controlling for the content creator, there are evidently other confounding factors that could account for some of the differences we found in this study, for instance user demographics or patterns of content discovery that are different across the two platforms.

In future work, we aim to expand and diversify our creator sample, map prevalent climate change discourse narratives on video-sharing platforms, and explore their correlation with audience reactions, with a focus on fostering commitment to action. 

\section{Related work} \label{sec:related}

In the realm of YouTube's climate change communication, there is a line of research examining the interplay between content and reactions to assess the effectiveness of communication strategies. In such a context, \citet{uldam2013online} explored YouTube's role in shaping communicative spaces for activism, \citet{shapiro2015more} analyzed the generality of reactions to climate change content and later on emphasized the presence of elite-driven discussions~\cite{shapiro2018climate}, while \citet{kaul2020environmental} provided insights on the role of influencers in engaging young audiences. These contributions are primarily qualitative and often involve the manual coding of videos.
In the case of TikTok, emphasis has been placed mainly on content rather than reactions. \citet{basch2022climate} studied the popularity of videos covering natural disasters, \citet{nguyen2023tiktok} used semantic networks and linguistic features for characterizing climate change videos, and \citet{hautea2021showing} presented an analysis of recurring templates in the video narratives of TikTok climate communication. Compared to previous literature, our study is the first that performs a cross-platform analysis between TikTok and YouTube and delves into the semantic relationship between content and reactions.

\begin{acks}
We acknowledge the support from the Carlsberg Foundation through the COCOONS project (CF21-0432)
\end{acks}

\balance

\bibliographystyle{ACM-Reference-Format}
\bibliography{sample-base}

\end{document}